\documentclass[a4paper]{article}
\usepackage{RRA4}
\usepackage{cite}
\usepackage{graphicx}
\usepackage{psfrag}
\usepackage{subfigure}
\usepackage{url}
\usepackage{stfloats}
\usepackage{amsmath,amssymb,array,verbatim,moreverb}
\usepackage{array}
\newtheorem{Lem}{Lemma}
\newtheorem{Thm}{Theorem}

\newtheorem{Cor}{Corollary}

\newenvironment{Proof}{\noindent Proof:
}{\hspace{\stretch{1}}$\square$}

 \RRdate{Ao\^{u}t 2006}
 \RRauthor{Julien Reynier \thanks{\'{E}cole Normale Sup\'{e}rieure, 45, rue d'Ulm, 75005 Paris}}
 \RRtitle{A simple stability condition for RED using TCP mean field modeling}
 \RRetitle{A simple stability condition for RED using TCP mean field modeling}
 \titlehead{Stability condition for RED}

\RRnote{TREC}
\RRabstract{
Congestion on the Internet is an old problem but still a subject of intensive research. The TCP protocol with its AIMD (Additive Increase and
Multiplicative Decrease) behavior hides very challenging problems; one of them is to understand the interaction between a large number of users with
delayed feedback.

This article will focus on two modeling issues of TCP which appeared to be important to tackle concrete scenarios when implementing the model
proposed in \cite{BMcDR02}; firstly the modeling of the maximum TCP window size: this maximum can be reached quickly in many practical cases; secondly the
delay structure: the usual Little-like formula behaves really poorly when queuing delays are variable, and may change dramatically the evolution of the
predicted queue size, which makes it useless to study drop-tail or RED (Random Early Detection) mechanisms.

Within proposed TCP modeling improvements, we are enabled to look at a concrete example where RED should be used in FIFO routers instead of letting
the default drop-tail happen. We study mathematically fixed points of the window size distribution and local stability of RED. An interesting case is
when RED operates at the limit when the congestion starts, it avoids unwanted loss of bandwidth and delay variations.
}

\RRkeyword{TCP, AQM, drop-tail, RED, congestion}

\RRresume{ Le contr\^{o}le de congestion dans Internet est depuis longtemps le sujet de recherches pouss\'{e}es. Le protocole TCP avec son comportement AIMD
(pour accroissements lin\'{e}aire, d\'{e}croissance multiplicative en anglais) cache des probl\`{e}mes excessivement compliqu\'{e}s. L'un d'entre eux est de
comprendre l'interaction entre de nombreux utilisateurs avec un d\'{e}lai de r\'{e}ponse du syst\`{e}me.

Ce rapport va se focaliser sur deux points dans la mod\'{e}lisation de TCP. Ces points sont apparus important lorsque nous avons voulu confronter \`{a} des
sc\'{e}narii concrets le mod\`{e}le propos\'{e} dans \cite{BMcDR02}; Tout d'abord la mod\'{e}lisation de la fen\^{e}tre maximale de TCP: cette valeur peur \^{e}tre atteinte
tr\`{e}s facilement dans la pratique; ensuite, la structure des d\'{e}lais: la formule type Little habituellement employ\'{e}e donne des r\'{e}sultats loin de la
r\'{e}alit\'{e} quand les d\'{e}lais sont variables. Cette hypoth\`{e}se de mod\'{e}lisation a un impact important sur la taille pr\'{e}dite de la file d'attente ce qui rend
vaines les tentatives de comparaison entre les m\'{e}canismes drop-tail et RED.

Gr\^{a}ce \`{a} ces am\'{e}liorations apport\'{e}es au mod\`{e}le, nous sommes capables dans un cas pr\'{e}cis d'\'{e}tudier comment param\'{e}trer RED dans des routeurs FIFO pour
qu'il am\'{e}liore les performances par rapport au cas par d\'{e}faut de drop-tail. Nous \'{e}tudions math\'{e}matiquement les points fixes et la distribution de la
taille des fen\^{e}tres et la stabilit\'{e} locale de RED. Un cas int\'{e}ressant est quand RED se trouve dans son domaine de fonctionnement au d\'{e}but de la
congestion, il \'{e}vite une mauvaise utilisation de la bande passante et des variations dans les d\'{e}lais. }

\RRmotcle{TCP, AQM, drop-tail, RED, congestion}
\RRprojets{TREC}
\RRtheme{\THCom}
\URRocq 
\begin{document}
\makeRR   

\tableofcontents

\section{Introduction}

\subsection{TCP router control issue}
TCP achieves a distributed congestion control of the Internet. This article proves a usable closed formula RED stability. RED (Random Early
Detection) was introduced by Floyd in \cite{floyd}; it is to be deployed at a router to send congestion information to TCP Reno end users. The idea
behind RED is that the first sign of congestion is when the router queue starts to be used more than to buffer normal traffic fluctuations; then the
buffer gets full and the drop-tail mechanism destroys packets arriving without any room left in the queue to fit in.

Drop-tail leads to two issues: firstly, the queue size oscillations provoke delay jitters - this has detrimental effects for applications using TCP
for realtime content - secondly, drop-tail synchronizes sources, resulting in bandwidth under-utilization of the congested link (this idea was first
introduced in \cite{shenker90some} for TCP Tahoe). This serves as leverage because at the time bandwidth demand reaches capacity, the goodput
diminishes by the synchronization effect, worsening the starting congestion (this assertion will be explained clearly later).

These two main reasons explain the interest for RED and other AQM (Active Queue Management) to deal with TCP congestion at the router level. RED
often works in an admirable way, leading to reduced queuing delays, avoiding jitters and reaching optimal bandwidth utilization... but sometimes RED
performs worse than doing nothing at all (drop-tail). This is the reason why many system administrators are reluctant to use RED although it is
deployed in almost every router of the Internet. This paper will show how to tune RED in a way it is sometimes optimal and always better than
drop-tail.

\subsection{Our previous works and motivation} In \cite{BMcDR02,McDR06} and \cite{Reynier06} we investigated mean field TCP modeling by continuing the
fluid TCP model introduced and studied in \cite{MGT00,hollot,liu03fluid}. Despite the interesting results arising from the models, there were still
some difficulties in understanding the original problem of tuning RED and comparing it accurately to drop-tail. Two points needed to be addressed.
Firstly, with the development of high speed access, it becomes difficult to suppose that TCP always works within its congestion avoidance mode in a
AIMD manner. The size of packets of the order of $1kB$ makes the maximum TCP window size relatively small (most common packet size is around $1.4
kB$). We shall say $W_{max}=64$ packets even if the receiver does not impose any reception window limitation. This fact is due to the coding of the
window size on $16$ bits addressing window by Bytes ($2^{16}B=64kB$).

Secondly, another limiting modeling assumption is a fact noticed by Hong in \cite{Hong_Fluid}: when the queue is not empty, acknowledgements arrive
obviously at the congested router bandwidth. This remark is crucial because TCP dynamic is very sensitive to the delayed feedback.

\subsection{Outline}
Section \ref{sec:Model} explains and defines our model; then we study the steady state window distribution with a maximal window size in section
\ref{sec:Fixed_Point}. Next step consists in seeking a stability region for the RED algorithm, which is done in section \ref{sec:Stability}. We
finish with showing simulation results on a concrete example in section \ref{sec:Simu}. This last example shows how to use previous results to
configure RED in a router in order to avoid collapse at the early stages of congestion.

\subsection{New results}
Whereas modeling $W_{max}$ and ACK bandwidth are not new ideas (see \cite{makowski-many} and \cite{Hong_Fluid}), adapting them in mean field
equations to obtain accurate evolution equations together with the window distribution constitutes a step forward. The steady state solution for the
window distribution taking into account the $W_{max}$ phenomenon in section \ref{sec:Fixed_Point} is an extension of \cite{BMcDR02} which is
important from a practical point of view. In section \ref{sec:Stability}, the stability result obtained for RED with theorem \ref{Thm:StableRED} is a
very simple closed formula. Finally the example in section \ref{sec:Simu} explains how RED should be tuned to increase router efficiency, this is an
important result because, as we said, the suggested tuning can be applied without any hardware modification in almost every router by enabling RED.

\section{Model and equations}\label{sec:Model}
\begin{figure}[h]
\includegraphics[width=\linewidth]{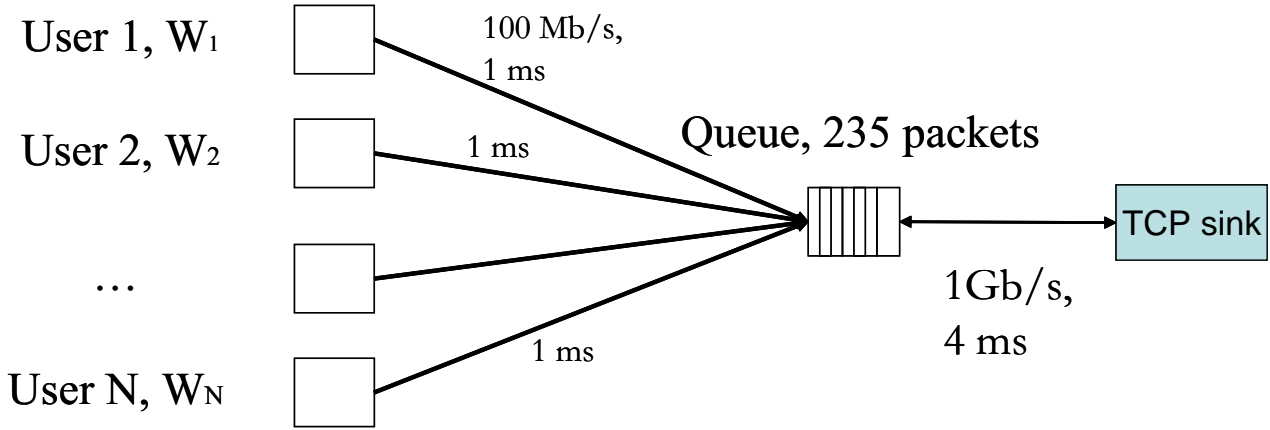}
\caption{Modeled topology and simulated scenario, $N$ is variable.} \label{Fig:TestBed}
\end{figure}

A number $N$, relatively large, of users share a common bottleneck router (figure \ref{Fig:TestBed} to see the modeled network topology). We can
consider the histogram of users' congestion window sizes; in \cite{McDR06}, we saw that this histogram converges "gently" to a deterministic window
distribution as $N$ tends to infinity. In \cite{BMcDR02}, we studied this asymptotic distribution which satisfies a partial differential equation the
results were applicable even for small numbers ($N=25$ for RED, and $N=10$ for a drop-tail). Hereinafter we adapt the partial differential equations
first presented in \cite{BMcDR02}; this is done in a way we could prove the mean field limit as we did in \cite{McDR06}, but we shall not enter in
such developments in this article.

\subsection{Evolution of congestion window sizes}
Imagine users have a notification of losses of the form $\kappa(t)$ in proportion of the incoming acknowledgement flow. Denote by $A(t)$ a function
indicating the flow evolution of windows (for example $A(t)=\frac{1}{rtt(t)}$ in usual TCP models). The question of delays and how the functions $A$
and $\kappa$ evolve come later in the article; these questions are not relevant to study the intrinsic user congestion window size evolution. Then,
the distribution of window sizes is of the form
$$D(t,w)=p(t,w)dw+M(t)\delta_{W_{max}}.$$ Which leads to two equations, the PDE:

\begin{align}\label{EDP}
&\frac{1}{A(t)}\frac{\partial p}{\partial t}(t,w)+\frac{\partial p}{\partial w}(t,w)= \\
&\ \ \kappa(t) \left( 4 w p(t,2w)\chi_{w<W_{max}/2}- w p(t,w)\chi_{w<W_{max}}\right)\nonumber\\
&\ \ +\delta_{\frac{W_{max}}{2}} M(t) \kappa(t) W_{max} \nonumber
\end{align}
and
\begin{equation}\label{EDP_W_max}
\frac{1}{A(t)}\frac{dM}{dt}(t)=p(t,W_{max})-M(t) \kappa(t) W_{max}.
\end{equation}

Intuitively, the coefficient $A(t)$ is the incoming bandwidth; when it is small, the window sizes have a slow reaction, when it is large, they react
in a faster way. The coefficients $\chi_{w<W_{max}}$ only indicate that the window size cannot be larger than $W_{max}$. When no loss occurs, the
coefficient $\frac{\partial p}{\partial w}(t,w)$ indicates that the window size increases linearly. When losses arise, the coefficient $\kappa$
enables $-\kappa w(p(w))$, which means that a certain proportion of users that were at window $w$ change to another value of the window size; it also
enables $4 \kappa w p(t,2w)$, which means that users that were at window $2w$ and $2w+1$ (or $2w-1$) move to window $w$.

\subsection{Delay in the system}
\subsubsection{Limits of Little-like formula} As noticed in \cite{Hong_Fluid} and in \cite{Reynier06}, the Little-like approximation made in usual
TCP models (for example \cite{hollot,BMcDR02,tina}) lacks realism and strongly limits the way models can explain reality. This approximation consists
in saying that at time $t$, the bandwidth $B(t)$ of a user is related to the RTT, $R(t)$ (Round-Trip Time) and its congestion window size $W(t)$ and
by $B(t)=W(t)/R(t)$. If the RTT is almost constant (for instance close to the propagation delay), it is a rather acceptable simplification, whereas
when $R(t)$ is variable, the model can lead to unacceptable consequences: it is easy to understand that when one wants to study the stability of RED
(with a non empty queue), saying $B(t)=W(t)/R(t)$ or $B(t)$ is constant entails different conclusions.

\subsubsection{How to improve delay model}
The idea comes from \cite{Bain} where a simple delay line is introduced to study the limit behavior (when the bandwidth tends to infinity) of one
user implementing MulTCP or scalable TCP (\cite{crowcroft98MulTCP,Kelly02,TomKelly04}). Although the use of a delay line complicates equations, the
model is still easy to simulate. Furthermore local stability of fixed points can be studied mathematically. Here we will adapt delay line modeling to
large number of TCP Reno users.

\subsubsection{Delay equations}
\paragraph{Delay and queue size}
Let us introduce $Q(t)$ the queue size mesured in seconds, the router is supposed FIFO (in other words, $Q(t)$ is the queuing delay). Denote by
$K(t)$, the destruction probability for a packet entering the queue; call $B_i(t)$ the incoming bandwidth to the queue and $B_o(t)$ the outgoing
bandwidth, scaled by the number of users. $C$ is the router capacity per user. Then:

\begin{equation}\label{EDP_Bo_Bi}
B_o(t)=\left\{  \begin{array}{c}
                  min(C,B_i(t)) \hbox{ if $Q(t)=0$}\\
                  C \hbox{ else.}\\
                \end{array}
\right.
\end{equation}

\paragraph{RTT}
Call $R(t)=T+Q(\tau(t))$, the RTT virtually written by the queue on packet arriving the router $\tau(t)$. This packet becomes an ACK that generates
new packets where the value is copied. By definition we say that this value comes back at the router router at time $t$, Which makes $t=\tau(t)+R(t)$
leading to the relation:
\begin{equation}\label{EDP_RTT}
R(t)=T+Q(t-R(t)).
\end{equation}
We discussed in \cite{BMcDR02} the fact that this implicit equation can also be written: $R(t+T+Q(t))=T+Q(t),$ which does not raise any definition
issues and is easier for numerical computations.

\paragraph{Advance $A(t)$ of window sizes}
The window size approximately increases by one every $W(t)$ arrived packets. The incoming bandwidth for a given user is the probability that the
packet is one of his multiplied by the total bandwidth:
$$\frac{W_j(t-R(t))}{\sum_i^N W_i(t-R(t))} B_o(t-T).$$

In fact we want to compute the advance of window size, which means that destroyed non-arriving packets carry information. Thus the modified ACK
bandwidth is:
$$\frac{1}{1-K(t-R(t))}\frac{W_j(t-R(t))}{\sum_i^N W_i(t-R(t))} B_o(t-T).$$
The factor of advance we shall use in window sizes evolution equations is:
\begin{equation}\label{EDP_A}
A(t)=\frac{1}{1-K(t-R(t))}\frac{1}{F(t-R(t))} B_o(t-T),
\end{equation}
where $F(t)=\frac{1}{N}\sum_i W_i(t)=\int w p(t,w)dw$ represents the number of packets on-the-flight.

\paragraph{Loss rate indicator} It is given by:
\begin{equation}\label{EDP_Kappa}
\kappa(t)=K(t-R(t)).
\end{equation}

\subsubsection{Bandwidth evolution when crossing the receiving user}
To go full circle\footnote{AKA: give the last equation} we need to say what is the value of the bandwidth $B_i(t)$ knowing the window size evolutions
and the ACK bandwidth $B_o(t)$. The evolution of this number only comes from new packets being sent or ACK being received (counting as ACK the
indication of a lost packet); thus:
\begin{equation}\label{EDP_Flight}
\frac{dF}{dt}(t)=B_i(t)-\frac{1}{1-K(t-R(t))}B_o(t-T).
\end{equation}

\subsubsection{Generation of losses}
We will suppose that losses are generated by some AQM (Active Queue Management), or by letting the drop-tail mechanism work. The equations are
(recall that $Q$ is given is seconds):
\begin{eqnarray}
\frac{dQ}{dt}=B_i(t)(1-K(t))-B_o(t) \label{EDP_Q}\\
\hbox{and } K(t)=f(B_i,Q).\nonumber
\end{eqnarray}
In the drop-tail case, for example, $K(t)=\frac{B_i(t)-B_o(t)}{B_i(t)}\chi_{Q(t)=Q_{max}}.$ For RED, with a loss function $f(\tilde{Q})$ and an
averaging coefficient $w_q$, $K(t)=f(\lambda\int_{-\infty}^t e^{\lambda(s-t)} Q(s)ds)$ with $\lambda=-N B_i\log(1-w_q)$. Hereinafter we shall suppose
the router uses RED with $\lambda$ very large which means that
\begin{equation}\label{EDP_AQM}
K(t)=f(Q(t)).
\end{equation}
To achieve this we shall say in the following that $w_q=1$, but a weaker assumption is that $\lambda>>1$, which would allow us to set $w_q$ as so to
admit bursts of packets without any losses when the total bandwidth is relatively large (see \cite{floyd}).

The model is completely specified by equations
(\ref{EDP},\ref{EDP_W_max},\ref{EDP_Bo_Bi},\ref{EDP_A},\ref{EDP_Kappa},\ref{EDP_Flight},\ref{EDP_Q},\ref{EDP_AQM}) and we can now analyse stability.
We first find fixed points with a constant loss indication, then we study the stability of these points supposing the use of a RED mechanism.

\section{Fixed point}\label{sec:Fixed_Point}
To study fixed points it is sufficient to study fixed points for window sizes. Other conditions follow immediately in paragraph
\ref{secsub:equilibrium_values}.
\subsection{Fixed point equations}
Eliminating $t$ in equations (\ref{EDP}) and (\ref{EDP_W_max}) leads us to consider a distribution of window sizes of the form:
$$D(w)=p(w)dw+M\delta_{W_{max}},$$ with the two equations:

\begin{eqnarray}\label{EDO}
p'(w)&=&4 k w p(2w)\chi_{w<W_{max}/2}-k w p(w)\chi_{w<W_{max}}\\
&&+M k W_{max}\delta_{\frac{W_{max}}{2}},\nonumber
\end{eqnarray}
and
\begin{equation}\label{EDO_Wmax}
M k W_{max}=p(W_{max}).
\end{equation}

\begin{figure}[h]
\includegraphics[width=\linewidth]{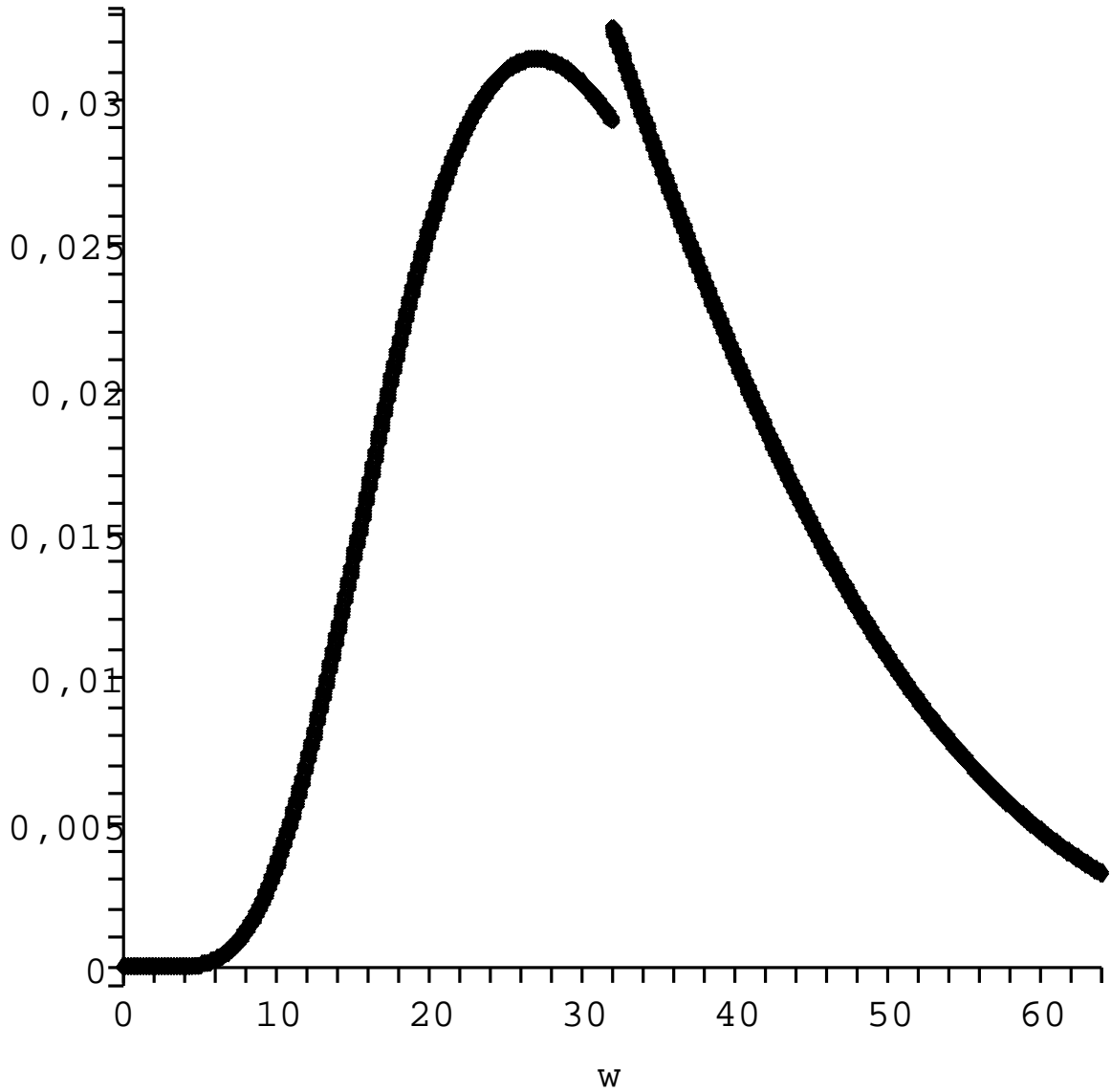}
\caption{p(w) for $W_{max}=64$, with the constant drop probability $0.15\%$, the mass at $W_{max}$ is approximately $3.3\%$.} \label{Fig:k_1_5p1000}
\end{figure}

The graphical representation of the solution found by MAPLE can be seen on figure \ref{Fig:k_1_5p1000}.

\subsection{Fixed point resolution}
\begin{Thm} The solutions of the system (\ref{EDO},\ref{EDO_Wmax}) on $\left[\frac{W_{max}}{2^{n+1}},\frac{W_{max}}{2^{n}}\right]$ are:
$$p(w)=\sum_{i=0}^n a_i^n e^{-4^i \frac{k w^2}{2}},$$
where for $n\geq 1$: $$a^n_0=a_0^{n-1} + M k W_{max} \frac{4^n}{\prod_{i=1}^n \left(4^i-1\right)} e^{-\frac{1}{4^n}\frac{k W_{max}^2}{2}}$$ and if
$i>0$: $a_i^n=\frac{4^{i}}{\prod_{l=1}^{i} \left(1-4^l\right) }a_{0}^{n-i}.$ One and only one solution is positive with integral $1$.
\end{Thm}

Given the formula, the theorem is simply verified by replacing the candidate solution in the equations. To see the intuition behind let us see the
first two iterations.
\subsubsection{First iteration} For
$w\in\left[\frac{W_{max}}{2},W_{max}\right]$, $$p'(w)=-k w p(w),$$ thus $p(w)=a_0^0 e^{-\frac{k w^2}{2}}$. The limit condition says that
$p(W_{max})=a_0^0 e^{-\frac{k W_{max}^2}{2}}=M k W_{max}$ which entails: $$a_0^0 = M k W_{max} e^{\frac{k W_{max}^2}{2}}.$$

\subsubsection{Second iteration} For $w\in\left[\frac{W_{max}}{4},\frac{W_{max}}{2}\right]$, $$p'(w)=-k w p(w)+4 k w a_0^0 e^{-4\frac{k w^2}{2}}.$$
By standard techniques we find the solution $p(w)=a_1^0 e^{-\frac{k w^2}{2}}-\frac{4}{3}a_0^0 e^{-4\frac{k w^2}{2}}$, which means that
$a^1_1:=-\frac{4}{3}a_0^0$. The limit condition determines $a_0^1$ by saying that
$p\left(\frac{W_{max}}{2}^+\right)=p\left(\frac{W_{max}}{2}^-\right)+M k W_{max}$, ie: $$a_0^0 e^{-\frac{1}{4}\frac{k W_{max}^2}{2}}=a_1^0
e^{-\frac{1}{4}\frac{k W_{max}^2}{2}}-\frac{4}{3}a_0^0 e^{-\frac{k W_{max}^2}{2}}+M k W_{max},$$

giving the good value to take for $a_1^0$: $a_1^0:= a_0^0 +MkW_{max} \frac{1}{4-1}e^{-\frac{1}{4}\frac{k W_{max}^2}{2}}$.

\subsection{Normalization}

\subsubsection{Regularity properties} We know {\it a priori} that for a well chosen value of the parameter $M$, $D$ is a probability. In this section, we
show a little more by saying that the density part $p$ is continuous and has a $0$ limit at $w=0$.

Notice that by integrating the EDO (\ref{EDO}) on $\left[ w, W_{max}+1 \right]$ we obtain
\begin{equation}\label{Somme_2^2w}
p(w)-0=\int_w^{2w} k v D(v).
\end{equation} We saw the solution
on $\left[\frac{W_{max}}{2},W_{max}\right]$ which in particular is positive, thus $p$ cannot reach $0$ for positive values of $w$ (we already knew
this by the fact that $D$ has to be a probability). By construction $p$ is continuous on $\left(0,\frac{W_{max}}{2}\right)$, thus bounded on compact
sets. Look at the explicit form of $p$ we have just calculated. It is always series bounded by:

\begin{eqnarray*}
\sum_{n=1}^\infty \frac{4^{i}}{\prod_{k=1}^{i}{4^i-1}} \left| a_0^n \right| &\leq& \sup_{k}\left| a_0^k \right| \sum_{n=1}^\infty
\frac{4^{i}}{\prod_{k=1}^{i}{4^i-1}}\\
&\leq& MW_{max}  \left(\sum_{n=1}^\infty \frac{4^{i}}{\prod_{k=1}^{i}{4^i-1}}\right)^2.
\end{eqnarray*}
And the last series is convergent because its general term is equivalent to $4^{-i (i-3)/2}$.

Now use the boundedness of $p$ in the equation (\ref{Somme_2^2w}) when $w$ is close to 0:
$$\left|p(w)\right|=\int_w^{2w} k v p(v)\leq (2w)^2k \sup \left|p\right|;$$
thus we see that $p$ tends towards $0$ at $w=0$.

\subsubsection{Computation of the integral}
\begin{figure}[h]
\includegraphics[width=\linewidth]{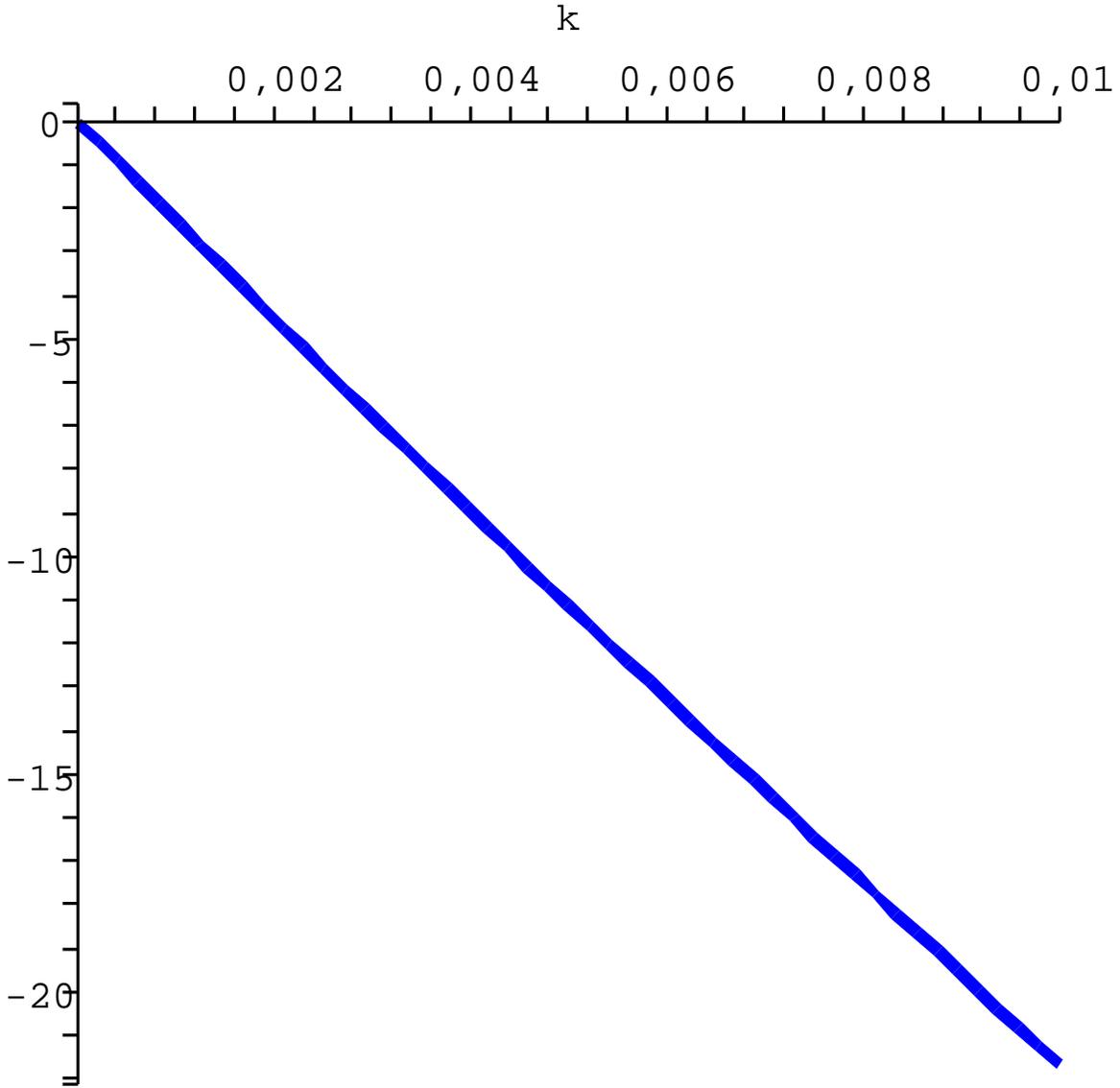}
\caption{$log(M)$ (the mass at $W_{max}$) function of the drop probability $k$ for $W_{max}=64$.} \label{Fig:log_M_k}
\end{figure}

\begin{eqnarray*}
\int_0^{W_{max}}D(v) = M + \sum_{n=0}^\infty \int_{W_{max}/2^{n+1}}^{W_{max}/2^{n}} \sum_{i=0}^n a_i^n e^{-4^i \frac{k v^2}{2}} dv\\
= M + \sum_{n=0}^\infty \sum_{i=0}^n a_i^n \sqrt{\frac{\pi}{2 k 4^i}} \vspace{-2cm}\times\left(erf(\sqrt{\frac{ 4^{i-n}}{2}k} W_{max})\right.\\
\ \ \ \left.-erf(\sqrt{\frac{4^{i-n}}{2\times 4} k} W_{max})\right).
\end{eqnarray*}
We already noticed that $p(w)\xrightarrow[w\to 0]{}0$, then it is not surprising that the previous sum converges very quickly by the conjugated
effects of $p$ being small and the size of the integration domain tending exponentially to $0$. Then a very interesting result from the practical
point of view is the proportion of users at $W_{max}$ function of the loss rate which is shown in figure \ref{Fig:log_M_k}. A first order Taylor
development of $log(M)$ in $k$ is immediate and:
\begin{Thm}\label{Thm:Taylor}
$$log(M)\sim -\frac{1}{2}k W_{max}^2.$$
\end{Thm}

\begin{figure}[h]
\includegraphics[width=\linewidth]{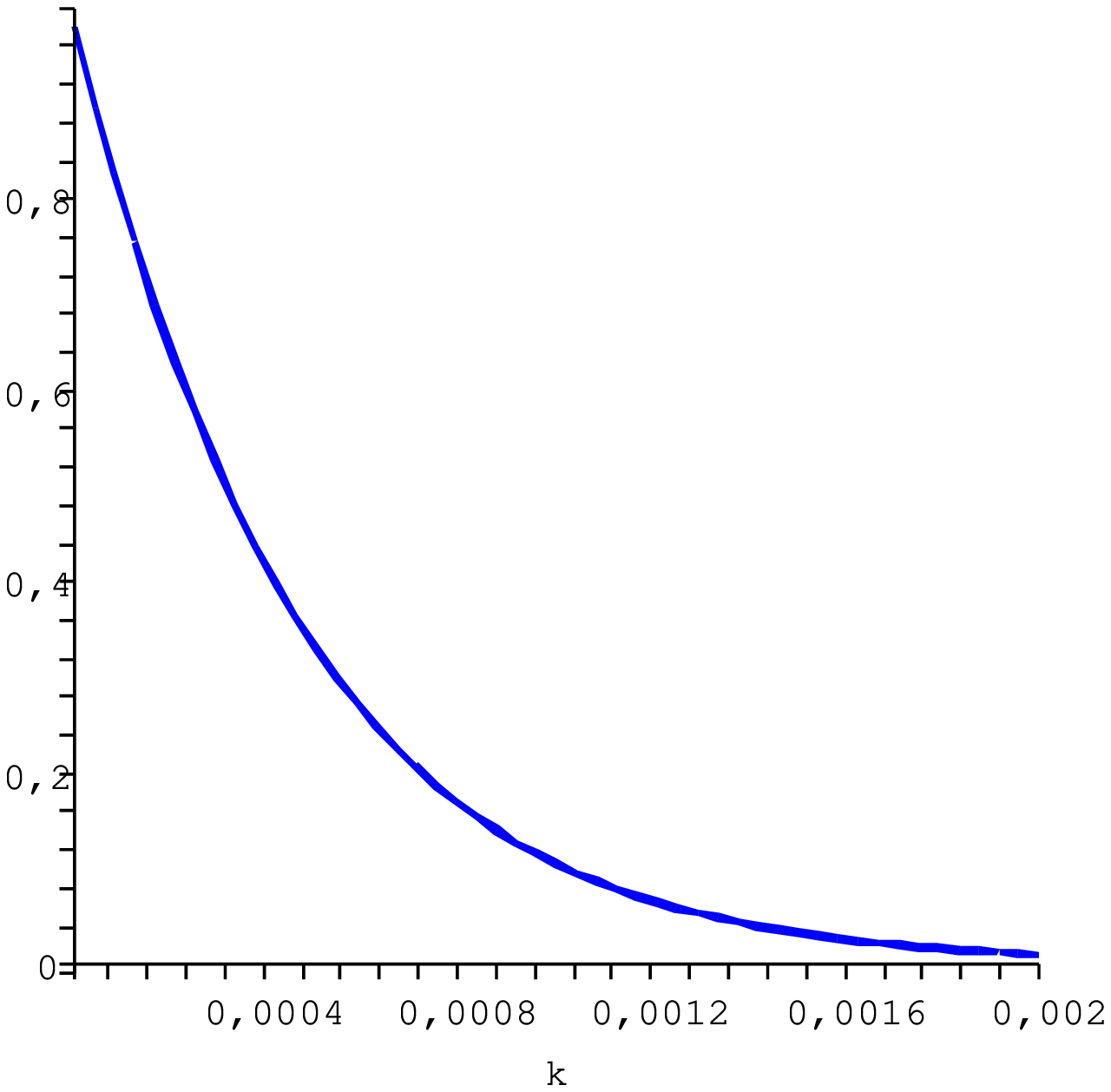}
\caption{$M$ (the mass at $W_{max}$) function of the drop probability $k$ for $W_{max}=64$.} \label{Fig:M_k}
\end{figure}

For instance see that the case of figure \ref{Fig:k_1_5p1000} gives approximately $M\approx e^{-0.15\% \times \frac{64^2}{2}}=4.6\%$, instead of the
exact value $3.3\%$. This is not a very accurate approximation, but it gives a good order of magnitude for a first approach, still, it is easy to
compute very good numerical values (see figure \ref{Fig:M_k}).

\subsection{Equilibrium values}\label{secsub:equilibrium_values}
Denote by $(B_i^e,B_o^e,K^e,Q^e,R^e)$ a set of equilibrium values with a non empty queue, then from equations (\ref{EDP_Bo_Bi}),(\ref{EDP_RTT}) and
(\ref{EDP_AQM}):

\begin{equation}
\left\{\begin{array}{ccc}\label{Eq_AQM}
        B^e_o&=&C\\
        R^e&=&T+Q^e\\
        K^e&=&F(B_i,Q^e).
\end{array} \right.
\end{equation}

At an equilibrium, the Little-like formula works (because the delay $R^e$ is constant) and:
\begin{equation}\label{Eq_B}
B_i^e=\frac{F^e}{R^e}.
\end{equation}

Then the conservation equation (\ref{EDP_Flight}) (or equation (\ref{EDP_Q})) and the advance equation (\ref{EDP_A}) give:
\begin{eqnarray}\label{Eq_Q}
B_i^e&=&\frac{C}{1-K^e}\\
\label{Eq_A} A^e&=&\frac{1}{R^e}.
\end{eqnarray}

This last equation shows us that in a steady state the function of advance $A$ is the one that usually appears in TCP models. We see that the
equilibrium relations are of the same kind as those in \cite{BMcDR02}.

\begin{figure}[h]
\includegraphics[width=\linewidth]{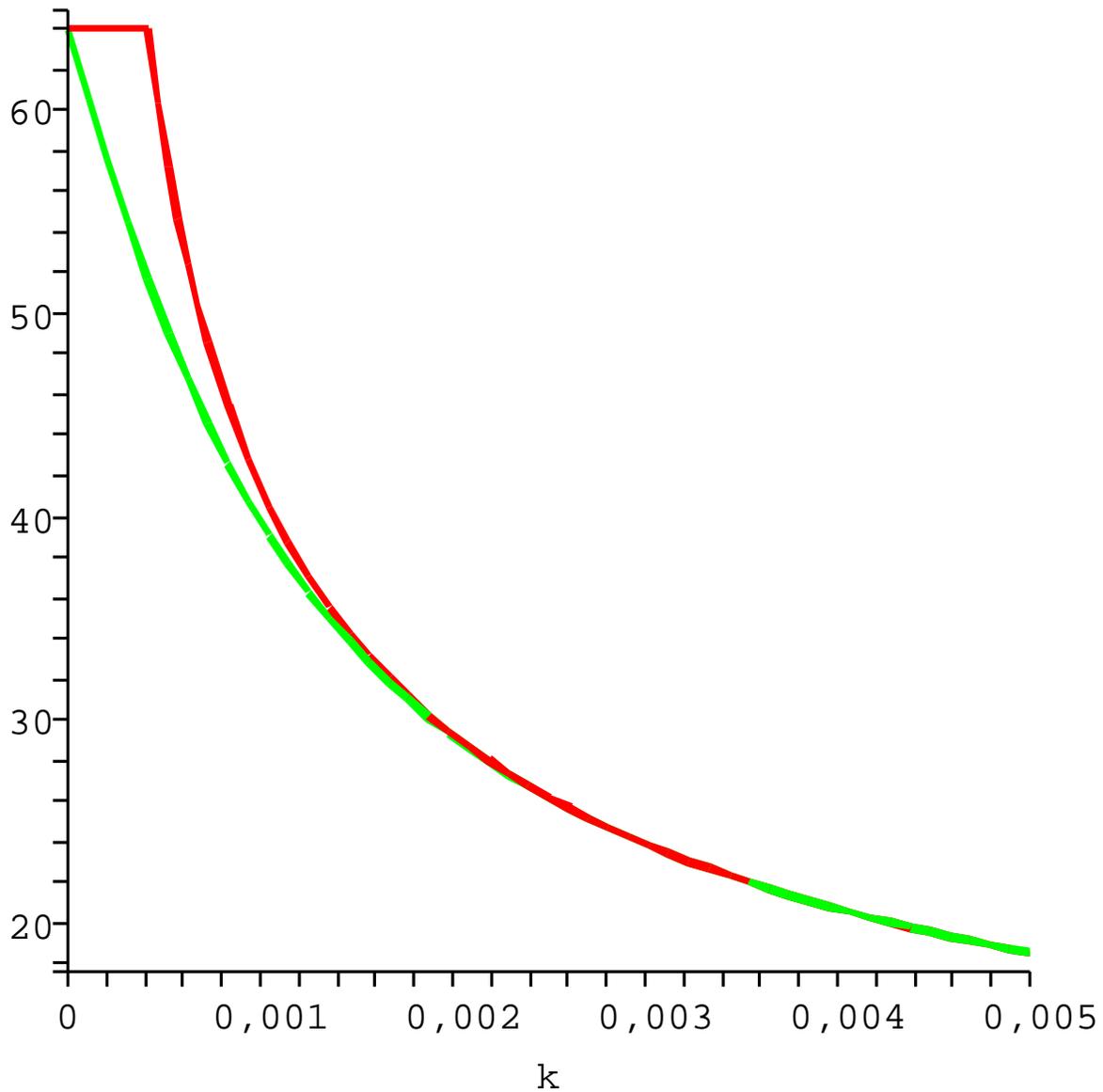}
\caption{Mean window size $F^e$ in light color compared to the square root formula from \cite{BMcDR02} cut at $W_{max}$:
$min(W_{max},\frac{1.31}{\sqrt{K^e}})$ in dark for $W_{max}=64$ and $K^e$ from $0$ to $0.5\%$.} \label{Fig:Squareroot}
\end{figure}

An equivalent to the square root formula would be needed; obviously when $K^e$ is not too small the square root formula still applies (the $W_{max}$
limitation is negligible), but when it starts to increase, the limitation on the window size lowers the mean window size. Figure \ref{Fig:Squareroot}
illustrates this result; we can see there that the square root formula is almost exact for $K^e>0.15\%$

\section{Stability analysis}\label{sec:Stability}
\subsection{A first remark}
Recall that $F(t)$ is the number of packets on the flight and let us call  $F_2(t):=\sum w^2 D(w)$ the second moment of the probability $D$. Denote
by a dot ($\dot{\square}$) the derivative of function $\square$ with respect to the time $t$. Then combining equations ($\int_w w \times $(\ref{EDP})
$dw + W_{max}$ (\ref{EDP_W_max})), we have:
\begin{equation}
\frac{1}{A(t)}\dot{F}(t)=1-M(t) - \frac{1}{2}K(t-R(t)) F_2(t),\label{EQ:Moments}
\end{equation}
which leads to the equilibrium equation:
\begin{equation}
\label{Eq_F_2} F_2^e=2\frac{1-M}{K}.
\end{equation}

\subsection{Stability equations}
We study the stability of the fixed point $(B_i^e,B_o^e,K^e,Q^e)$. We intend to study an equilibrium with a non-empty queue, this implies
$B_o(t)=B_o^e=C$.

The idea is to add a small perturbation of the form $\Delta w$ on the window sizes at $t=0$ a time at which a fixed point has been reached. To
simplify we suppose that the response is uniform and we denote it by $\Delta w(t)$; the variations are truncated at the first order. This
simplifications entails that the variation of the on the flight packets number $F(t)=\sum w D(w)$ is $\Delta F(t)=\Delta w(t)$.

The assumption on $\Delta w$ permits to write $\Delta F_2(t)= 2 F^e\Delta F$; then taking the variation at first order in equation (\ref{EQ:Moments})
gives:
\begin{equation}\label{Linear_Moments}
\Delta \dot{F}(t) = \left\{\begin{array}{c} A^e \Delta M(t)- A^e K^e F^e \Delta F\\
- \frac{1}{2} A^e F_2^e \Delta K(t-R^e).
\end{array}\right.
\end{equation}

This equation comes with the linearized version of (\ref{EDP_W_max}):
\begin{equation}\label{Linear_W_max}
\Delta \dot{M}(t) = \left\{\begin{array}{c}
A^e\Delta p(t,W_{max})- A^e K^e W_{max} \Delta M(t)\\
- A^e M^e W_{max} \Delta K(t-R^e).
\end{array}\right.
\end{equation}

From equation (\ref{EDP_Flight}):
\begin{equation}\label{Linear_Flight}
\Delta \dot{F}(t) = \Delta B_i(t)- \frac{C}{\left(1-K^e\right)^2} \Delta K(t-R^e).
\end{equation}

Equation (\ref{EDP_Q}) leads to:
\begin{equation}\label{Linear_Q}
C \Delta \dot{Q}(t) = (1-K^e) \Delta B_i(t) - B_i^e \Delta K(t).
\end{equation}

Finally the instantaneous RED control gives:
\begin{equation}\label{Linear_RED}
\Delta K(t) = \epsilon \Delta Q(t).
\end{equation}
where $\epsilon$ is the slope of the RED control function at the equilibrium point $Q^e$. We suppose here that the averaging factor $w_q$ is equal to
$1$ (which means that we only consider the instantaneous value of the queue to compute losses).

Notice that the variations of RTT only create second order terms; this is the reason why equation (\ref{EDP_RTT}) does not have to be used. This is
the same for equation (\ref{EDP_A}), because the fist term factors are always multiplied by the second term of the equations that have null
equilibrium values.

\subsection{Differential equations with time delay}
The equations (\ref{Linear_Moments}), (\ref{Linear_W_max}), (\ref{Linear_Flight}), (\ref{Linear_Q}), (\ref{Linear_RED}) can be reorganized as a
system of three delay differential equations on $F$, $Q$ and $M$. The only problem is that $\Delta p(W_{max},t)$ is not well determined in the
equation (\ref{Linear_Moments}). We shall make the further simplifying assumption that the term $A^e\Delta p(t,W_{max})$ can be replaced by
$p^e(W_{max})\Delta w=p^e(W_{max})\Delta F(t)$, which is intuitive since when all windows are increased by $\Delta w$, the additional number of users
overtaking window $W_{max}$ is close to the announced number if we say that the window distribution stays close to the equilibrium one at the first
order.

The standard mathematical method to find the local stability condition with a linear delay differential equation is the following (see
\cite{DelayDE}; this is an equivalent to the Bode diagrams approach). Find a solution of the form $\lambda_i e^{\phi t}$ where $\lambda_i$ and $\phi$
are complex numbers, linearize the exponential factor coming from the delayed terms which corresponds to the replacement of terms $\Delta X(t-r)$ by
$(1-r \phi)\Delta X(t)$. We are led to one more unknown than equations, hopefully one of the $\lambda$-s may be replaced by the value $1$. Then solve
the polynomial system to find $\phi$. A necessary condition for the system to be stable is that the real parts of all solutions are negative.

Let us look for a solution of the form $(\Delta F,\Delta M, \Delta Q)=(e^{\phi t}, x e^{\phi t}, y e^{ \phi t})$. This leads us to (the equations are
divided by $e^{\phi t}$ and the expressions $e^{\phi R^e}$ are linearized for $\phi R^e<<1$):
\begin{eqnarray}
R^e \phi&=&- K^e F^e - x -\frac{\epsilon(1-M^e)}{K}(1-R^e\phi)y\label{Sta1}\\
R^e \phi x &=& R^e M^e K^e W_{max} - K^e W_{max} x\nonumber\\
&&-M^e W_{max}\epsilon (1-R^e\phi) y\label{Sta2}\\
C\phi y &=& (1-K^e)\phi - \frac{R^e C\epsilon}{1-K^e}\phi y.\label{Sta3}
\end{eqnarray}

From (\ref{Sta3}) and replacing using (\ref{Eq_B}) and (\ref{Eq_Q}):
\begin{equation}\label{Val_y}
y=\frac{1-K^e}{C+\epsilon F^e}.
\end{equation}

Then from (\ref{Sta1}) using (\ref{Eq_A}):
\begin{equation}\label{Val_x}
-x=R^e \phi + K^eF^e+\frac{(1-M)\epsilon}{K}(1-R^e\phi)y.
\end{equation}

Putting everything in (\ref{Sta2}):

\begin{eqnarray}
(R^e\phi + K^e W_{max})\left(-x\right)+R^eM^e K^e W_{max}\nonumber\\
-M^e W_{max}\epsilon (1-R^e\phi) y&=&0.\label{Sta4}
\end{eqnarray}

This is a second degree equation of the form:
\begin{equation}\label{binome}
a\phi^2+b\phi+c=0,
\end{equation}
with:
$$\begin{array}{ccc}
a&:=&(R^e)^2(1-\frac{1-M^e}{K^e} \epsilon y)\\
b&:=& R^e\left[K^eF^e + \frac{\epsilon y}{K}(1-M^e)+K^eW_{max}\right.\\
&&\left.-K^eW_{max}\frac{\epsilon y}{K}(1-M^e)+M^eW_{max}\right]\\
c&:=& \hbox{not needed}.\end{array}$$
\begin{Lem}\label{Lem:binome}
The following properties are fulfilled:
\begin{itemize}
 \item $a,b$ and $c$ are real numbers,
 \item Suppose $a>0$, then both solutions have negative real values if and only if $b>0$.
\end{itemize}
\end{Lem}
\begin{Proof}
The first point is true by definition. The second point is easy: if $\phi_1$ and $\phi_2$ are the roots of (\ref{binome}), then
$\phi_1+\phi_2=-\frac{b}{a}$. The coefficients are real, which ensures that $\phi_2=\bar{\phi_1}$ (the conjugated complex number), thus
$\phi_1+\phi_2=2\mathcal{R}e(\phi_1)$, which grants our point.
\end{Proof}

\begin{Thm}\label{Thm:StableRED_petit}
A sufficient condition for RED with $w_q=1$ to be stable is:
\begin{equation}\label{StableRED}
\epsilon<\frac{K^e C}{1-M^e}.
\end{equation}
\end{Thm}
\begin{Proof}
First look at $b$; let $U=\frac{\epsilon y}{K^e}(1-M^e)>0$, then: $\frac{b}{R^e}>0$ if and only if:
$$K^eF^e+U + K W_{max}(1-U)+M^e W_{max}>0.$$
A sufficient condition is that $U<1$ but: $\epsilon<\frac{K^e C}{1-M^e}\Rightarrow \epsilon(1-M^e)(1-K) < K^e C$, which implies $\epsilon(1-M^e)(1-K)
< K^e C+K^e\epsilon F^e,$ ie: $U<1$.

$\frac{a}{(R^e)^2}=1-U>0$ like we have just seen. Then Lemma \ref{Lem:binome} applies and gives the conclusion.

To finish the proof, we need to say something about the assumption $\phi R^e<<1$. An acceptable condition would be that $\frac{b}{2a}R^e<<1$ (we only
check the that the real part is small):
$$\frac{b}{2a}R^e=\frac{1-U}{K^eF^e+U+W_{max}K^e(1-U)+M^eW_{max}}.$$
We see there that the RTT does not play an important role; this quantity is small if  either $K^eW_{max}$ or $M^eW_{max}$ is large, which means that
is is always a good approximation.
\end{Proof}

Remark that from the proof a weaker stability condition for RED with $w_q=1$ is $$\frac{1-K^e}{K^e}\frac{\epsilon}{C+\epsilon F^e}(1-M^e)=U<1.$$

\begin{Cor}
If $W_{max}=\infty$ (the approximation made in \cite{BMcDR02}), then the stability condition for RED with $w_q=1$ becomes in all the usual conditions
\footnote{for $K<54\%$ which is a lot larger than the limits tolerated by TCP that turn around $8\%$}:
$$\epsilon<\frac{KC}{1-\alpha \sqrt{K}-K}.$$
with $\alpha\approx 1.310$.
\end{Cor}
\begin{Proof}
The proof would be the same without the second equation on $M$. In that case we found in \cite{BMcDR02} the exact formula:
$F^e=\frac{\alpha}{\sqrt{K}}$ (this is one example of the well-known TCP square root formula). All this directly leads to:
$$\frac{1-K^e}{K^e}\frac{\epsilon}{C+\epsilon \frac{\alpha}{\sqrt{K}}}<1.$$
The conclusion is only a reorganization of this equation.
\end{Proof}

We see that that the condition $\epsilon<RC$ is a rule of the thumb valid in every case. The last corollary will be named theorem because it is the
most important result of the article from a technical point of view.

\begin{Thm}\label{Thm:StableRED}
A universal stability condition for RED is: $$\epsilon<\frac{\alpha^2}{(T+Q_{max}) W_{max}},$$ where $\alpha^2\approx 1.7$. For parameters $T$,
$Q_{max}=Max_{th}=\beta T$, $Min_{th}=\gamma T$, and $p_{max}$ with $w_q=1$, RED is stable if:
\begin{equation}
p_{max}<\frac{\beta-\gamma}{\beta+1}\frac{\alpha^2}{W_{max}}.
\end{equation}
\end{Thm}

\begin{Proof}
Recall theorem \ref{Thm:Taylor} says that $M^e\sim e^{-\frac{1}{2}K^e W_{max}^2}$ when $K^e$ is close to $0$, then: $\epsilon<\frac{2C}{W_{max}^2}$
is a stability condition; which entails the result for $K^e$ close to $0$, using the fact that the capacity for a user at the window $W_{max}$ is
exactly $\frac{W_{max}}{R^e}$. For other values, the square root formula implies that
$\frac{K}{1-K}C>\frac{\alpha^2}{(1-K)(R^e)^2C}>\frac{\alpha^2}{RW_{max}}$. To conclude, add the fact that $R^e<T+Q_{max}$ and the definition of RED.
\end{Proof}

\section{Simulation results}\label{sec:Simu}
\begin{figure}[h]
\includegraphics[width=\linewidth]{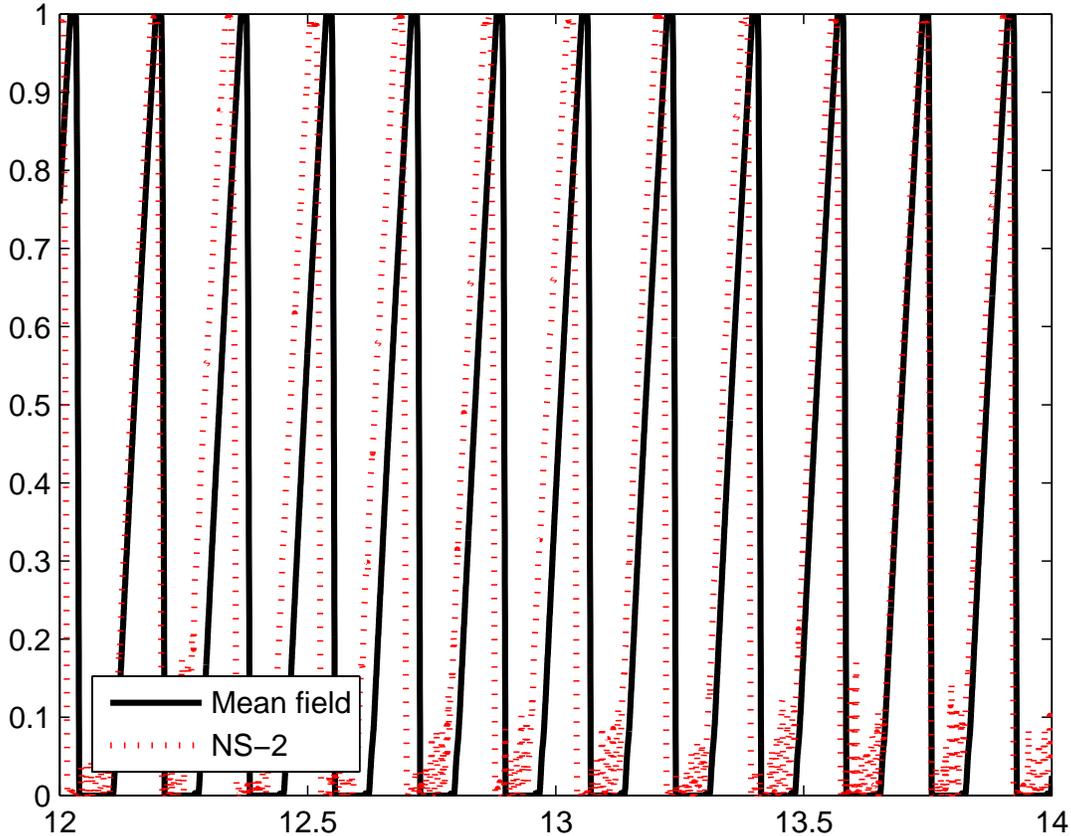}
\caption{Normalized queue size for $35$ users with $1kB$ packets on $1Gb/s$ link and drop-tail policy (time is in seconds, propagation delay 10 ms,
maximum queuing delay 2ms). NS-2 is dotted and mean field equations give the solid line.} \label{Fig:N35_Queue_DT_NSvsMatlab}
\end{figure}

The example we shall study is inspired by a real Internet provider configuration, it is illustrated by figure \ref{Fig:TestBed}; the mean field
simulator can be downloaded at \cite{HTTP_program}. On a one giga-bit router in some part of the network the total propagation delay for end users is
$10 ms$. The router is configured with a $2 ms$ FIFO buffer (which is five time less than the usual delay bandwidth product rule). The faced problem
is a jitter felt by end users. The size of packets is supposed to be $1kB=8192\ bits$ and we shall say that the level 2 overhead is $40B$; then the
maximum congestion window size which is $64 kB$ corresponds to $64$ packets; the router capacity is $1.17\hbox{e+}5$ packets per second and the
buffer size corresponds to $235$ packets. We also suppose that end users have a limited capacity at their access so that the packets do not arrive in
bursts at the router (which is an assumption of our loss model); let us say that the limit is $100Mbits/s$ (and the buffer size at the access is
unlimited).\\
In \cite{Reynier06}, we saw that $10$ users can be considered to be a large number for the drop-tail. When sources are less synchronized, the mean
field simulations are always accurate for $25$ or more users. We can see this on figure \ref{Fig:N35_Queue_DT_NSvsMatlab} that the NS simulation of
TCP Reno works close to our model which means that there are few timeouts and slow starts and that AIMD is a good model for {\it fast recovery/fast
retransmit}.

\subsection{Results with drop-tail}
\subsubsection{Before congestion happens} 
\begin{figure}[h]
\includegraphics[width=\linewidth]{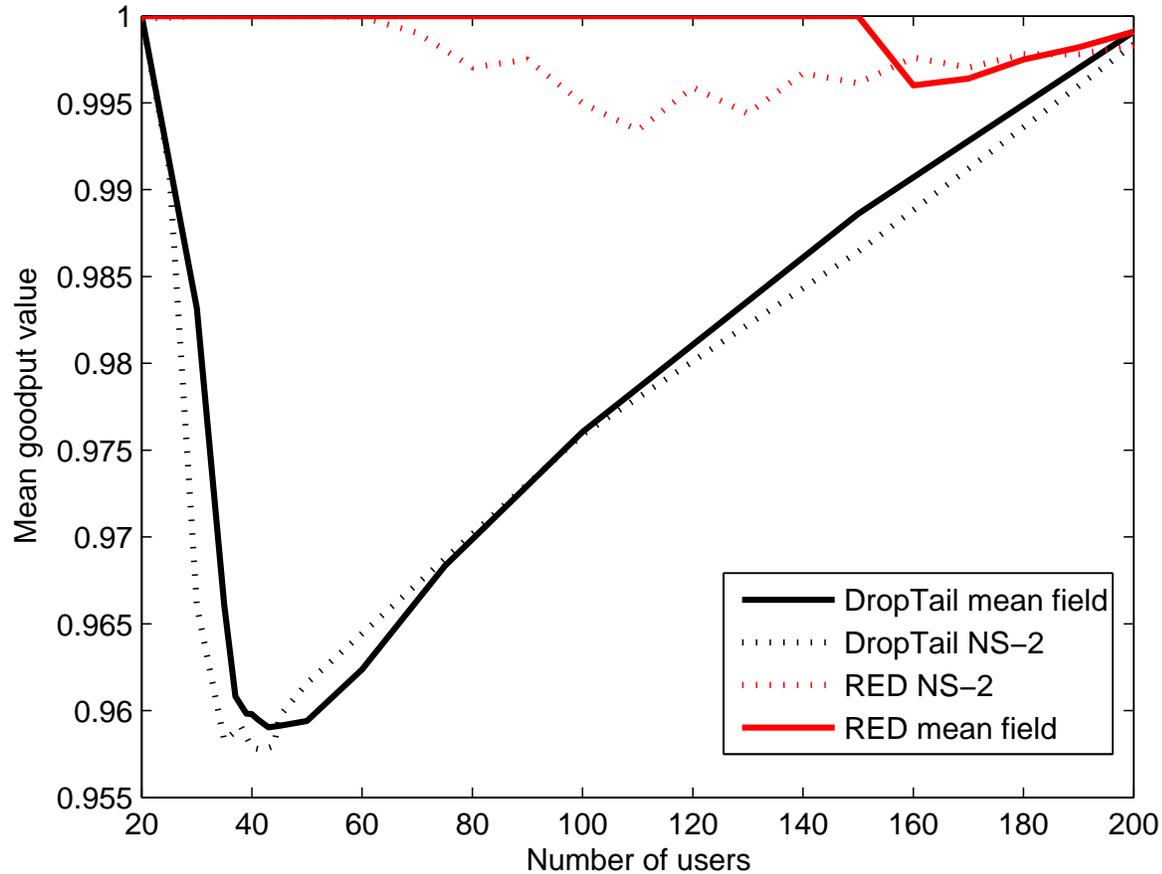}
\caption{Comparison between RED and DropTail for NS-2 and our mean field equations. The network model is the one of figure (\ref{Fig:TestBed}); RED
parameters are those of paragraph \ref{configuration}.} \label{Fig:RED_BW}
\end{figure}

As can be seen in figure \ref{Fig:RED_BW}, for less than $19$ users, the router capacity cannot be reached
and the total throughput per user stays at $\frac{W_{max}}{T}*1024*8 bits \approx 52 Mbits/s$ which is the maximum possible with the considered
propagation delay and packet size TCP can allow. We see that from $20$ to $22$ users, the queue increases steadily from $0$ to its maximal value, so
the RTT increases from $10ms$ to $12ms$. Remark that a stable queue close to its maximal value is something that should be avoided because it leaves
too little room for fluctuations to be smoothed. For 23 and 24 users, the queue starts oscillating, but the bandwidth still stays around its maximum.

\subsubsection{The early congestion phase}
From $25$ users, both NS-2 and the mean field equations show an extremely bad behavior: the utilization drops to $99\%$ for NS-2. Then utilization
drops to a worst utilization of $96\%$ around $40$ users, this can be explained by an increasing synchronization between users.

\subsubsection{Strong congestion phase} For more than $50$ users, the utilization starts to increase because the mean window size decreases: although
the synchronization level is very high, with a small window, the additive increase mechanism goes back to a maximal utilization quicker than with a
larger window which explains the link utilization improvement.

\subsection{Results with RED}
\subsubsection{Configuration}\label{configuration}
Suppose that Max$_{th}=Q_{max}=T/5$ and that Min$_{th}=0.2$ Max$_{th}=.4 ms=47$ packets. The rule of the thumb of theorem \ref{Thm:StableRED} gives a
value of $.36\%$, which gives an insight of the value to take. We saw with NS and by simulating the mean field equations that $.5\%$ was also a
working value whereas $.75\%$ was too high to achieve a stabilization in every case (but leads to small oscillations), which explains our choice:
$p_{max}:=.5\%.$

Nothing changes for less than $20$ users because the queue size stays below the minimum threshold of RED.
\subsubsection{RED in its working regime}
From $20$ to $80$ users, RED permits to have a steady state with a queue size going to its maximal value. From $21$ to $70$ users the queue size goes
from $0$ to $1ms$, the second half of the queue size is the stabilization region between $71$ and $80$ users. This can be explained roughly by the
square-root formula: the steady state value of the loss rate is proportional to $(TC)^{-2}$, when $C$ diminishes, the loss rate increases
quadratically and so does the queue size. This fact would advocate for an exponential shape of the loss rate function as indicated in
\cite{Reynier06}.
\subsubsection{RED working like a drop-tail}
Then, in the case $81$, the simulation noise in NS makes the queue size touch the border and begin a drop-tail like behavior. So does the mean field
simulator for $85$ users. Then RED behaves like an improved version of a drop-tail (for less than $130$ users the queue never empties). Overall
figure \ref{Fig:RED_BW} shows that the bandwidth utilization always stays beyond $99.5\%$. In this state RED behaves better than drop-tail from the
bandwidth utilization point of view, but there is an oscillation which makes it a good choice to take a small queue.

\subsection{Increasing the latency}
When one increases the latency, the relative value of $Q_{max}$ decreases, meaning that even with the same synchronization, the buffer does not
provide the same bandwidth insurance. Another effect has to be taken care of: when the latency increases, the maximal bandwidth decreases, which
means that more users are needed to reach the router capacity. When the latency is increased, the worst case for drop-tail is still at the early
congestion stage because window sizes are huge. We saw that our RED configuration, even when not working in the steady state domain, gives better
results in terms of link utilization.

\subsection{Mixing latencies}
When latencies are mixed, as was previously observed in the literature, the equivalent latency is the harmonic mean of latencies, meaning that small
latencies are preponderant in the configuration of a router. This fact is intuitive because the small latency connections adapt to bandwidth changes
quicker, and if they are stabilized by the controller, the set of other connections act exactly like one constant bitrate user (even if each one of
those connections sometimes divides its bandwidth by a factor 2). We also observed that when RED was not acting in its steady state area, our RED
configuration never acted in a worst way than drop-tail, which is due to the fact that $p_{max}$ is not too large. The case where RED would be worse
than drop-tail would be for a too large value of $p_{max}$ where RED acts like a drop-tail at Min$_{th}$ which means that a part of the buffer is
never used.

\section{Conclusion}\label{sec:Conclusion}
We saw how to model accurately TCP and how to give an easy closed formula to tune RED. This lead us to observe a bad news about the drop-tail: the
worst case for bandwidth utilization for a drop-tail is just after the congestion is reached. This is illustrated in our example. We saw there how to
use our framework to configure properly RED to obtain a situation where the congestion can be supported without any loss of bandwidth for a very long
time and without any delay oscillations. Then for extreme values, our configuration behaves not worse than drop-tail which is a good reason to use
RED in a router. In an actual router users have multiple latencies, we also said briefly that if a sufficient number of low latency connections are
present, then RED leads to a steady state.

\section{Related Works}\label{sec:Related_Works}
\subsection{TCP modeling area}
The problem of $N$ connections sharing one bottleneck router has been extensively studied in past years. The first models were made by Ott and Al. in
\cite{ott96stationary,mathis97macroscopic,OLW99a,CJOS00,MO02}. Then some interesting studies belong to May, Bonald and Bolot in \cite{MBB00} and
Vinnicombe in \cite{vin:ont02}, but it appeared we owe the most promising approaches to Kelly and Al. \cite{KMT98a,low:red00} with a utility
maximization problem and to Gong, Hollot, Misra and Towsley in \cite{MGT00,hollot,liu03fluid} with the idea of introducing a fluid equation supposed
to model the aggregated behavior of many TCP sources. This last approach motivated mathematical study of the mean field interaction to obtain
accurate intrinsic equations of what TCP is; namely it was the study of the AIMD TCP Reno behavior (congestion avoidance
\cite{jacobson88congestion}).

The main works in the area are those by Tinnakornsrisuphap and Makowski \cite{tina,makowski-many,tinnakornsrisuphap03limit} with a discrete time
model simple yet very efficient; Srikant and Al. \cite{Srikant,Srikant2} with discrete time where TCP users have to compete against a white noise;
Baccelli, Hong and Al. \cite{HL01,BHo02b,BHo03b,chaintreau02closed,baccelli03flow} with stochastic time steps, no buffer but an optional HTTP
adaptation \cite{baccelli04meanfield}; and Baccelli, McDonald and Reynier \cite{BMcDR02,McDR06} which is the model we adapted in this article.

We believe our model is the most efficient because we were able to use continuous times which really matters due to the strong dependence of the
problem on delay; our model explicitly uses the TCP mechanism and we were able to deal with boundary effects which made it possible to study both RED
(or other AQM mechanisms) and the drop-tail. We were also able to take into account heterogeneous sources (see \cite{McDR06}). This article permits
to see one other advantage of our approach, it is easily adaptable to changes in the TCP dynamic or in the way TCP is modeled; for example in
\cite{Reynier06}, we saw how to adapt it to intermittent TCP sources (to model HTTP users behavior).

\subsection{Control theory applied to TCP}
Another branch of studies is the control theoretic approach used in \cite{hollot} we adapted here to find stability conditions for the time delayed
equations we dealt with; for example, the same kind method was used by Kim and Low in \cite{klascc:aqm02}. The problem of these studies is that they
usually rely on a little-like formula, which leads to poor results when trying to compare to simulations: simulations show behaviors a lot nicer than
expected. Here we solved this issue and found a very simple closed formula that implies stability for RED (see theorem \ref{Thm:StableRED}).

\subsection{Buffer sizing for IP routers}
As noticed by McKeown, Wischik and Al. in \cite{virtamo,Wischik,Wischik2,Wischik3,Appenzeller}, the kind of scaling we do in our model can create
problems. In core routers, slowly switching from ATM to IP, very fast and expensive memory is needed, and bandwidth optimization is not the first
goal. In that case good overall performances can be achieved by choosing very small buffers at the cost of a waste of bandwidth even before the
congestion level is reached. We did not intent to study highspeed core routers in this article. We are interested in some access routers that are not
in the provider's backbone. The bandwidth is limited and the number of links to upgrade make it difficult to over provision users' needs. Then, as we
saw in the simulation section, RED may be a solution to avoid the leverage effect at the early stages of congestion.

\section{Further Works}\label{sec:Further_Works} Understanding exactly how to tune a router to avoid early congestion effects for HTTP users is
still a challenge. Even if the equations are relatively easy to write (see \cite{Reynier06} or \cite{baccelli04squareroot} for theory and the
implementation in \cite{HTTP_program}), from a practical point of view it is difficult to obtain accurate results. This is because of a high output
dependence on how users are modeled, and from their statistics. For instance, determining what is a "good" distribution of file sizes or idle times
between two downloads is not an easy task.

Another interesting task would be to obtain easy closed formulae for drop-tail metrics such as bandwidth utilization.

\section*{Acknowledgments}
The author would like to thank Thomas Bonald, Dohy Hong, Fran\c{c}ois Baccelli, David McDonald, Ki-Beak Kim for their kind careful help and thorough
suggestions. A special thank to Anamaria for her rereading.

\bibliographystyle{amsplain}
\bibliography{IEEEfull,Bibliographie}

\end{document}